\documentclass[a4paper,11pt]{article}
\usepackage{aaskaiid}
\usepackage{threeparttable}
\usepackage{orcidlink}
\title{Mapping the Milky Way with Masers}
\ShortTitle{MW Spiral Structure}

\author[1]{$^{\ast}$Ye Xu\orcidlink{0000-0001-5602-3306}}
\ShortName{Xu et al.} 
\author[2]{Kazi Rygl\orcidlink{0000-0003-4146-9043}}
\author[3,4]{Huib Jan van Langevelde\orcidlink{0000-0002-0230-5946}}
\author[5,6]{Dejian Liu\orcidlink{0009-0001-9837-9455}}
\author[1]{Jingjing Li\orcidlink{0000-0002-3338-8465}}
\author[1]{Yingjie Li\orcidlink{0000-0001-7526-0120}}
\author[7,8]{Simon P. Ellingsen\orcidlink{0000-0002-1363-5457}}
\author[8,9]{Mar\'{i}a J. Rioja\orcidlink{0000-0003-4871-9535}}
\author[8]{Richard Dodson\orcidlink{0000-0003-0392-3604}}
\author[1]{Zehao Lin\orcidlink{0009-0006-0392-6345}}
\author[1]{Chaojie Hao\orcidlink{0000-0002-6820-198X}}
\author[1,10]{Yiwei Dong\orcidlink{0009-0000-5512-9109}}
\author[11]{Jun Yang\orcidlink{0000-0002-2322-5232}}

\affiliation[1]{Purple Mountain Observatory, Chinese Academy of Sciences, Nanjing 210023, China}
\affiliation[2]{INAF-Istituto di Radioastronomia \& Italian ALMA Regional Centre, Italy}
\affiliation[3]{Joint Institute for VLBI ERIC (JIVE), Netherlands, Oude Hoogeveensedijk 4, 7991 PD Dwingeloo}
\affiliation[4]{Leiden Observatory, Leiden University, PO Box 9513, 2300 RA Leiden, the Netherlands}
\affiliation[5]{College of Science, China Three Gorges University, Yichang 443002, China}
\affiliation[6]{Center for Astronomy and Space Sciences, China Three Gorges University, Yichang 443002, China}
\affiliation[7]{School of Natural Sciences, University of Tasmania, Private Bag 37, Hobart, Tasmania 7001, Australia}
\affiliation[8]{ International Centre for Radio Astronomy Research, The University of Western Australia, 35 Stirling Hwy, Crawley, WA, Australia}
\affiliation[9]{Observatorio Astron\'{o}mico Nacional (IGN), Alfonso XII, 3 y 5, 28014 Madrid, Spain}
\affiliation[10]{School of Astronomy and Space Science, University of Science and Technology of China, Hefei 230026, China}
\affiliation[11]{Department of Space, Earth and Environment, Chalmers University of Technology, Onsala Space Observatory, 43992 Onsala, Sweden}
\emailAdd{xuye@pmo.ac.cn}

\abstract{SKA-VLBI is poised to revolutionize our understanding of the Galactic structure through its unprecedented astrometric precision and sensitivity. As a next-generation facility, it will answer long-standing questions about the Galactic structure by mapping its entire spiral structure in detail, spanning from the solar neighborhood, through the Galactic Center, to the far side of the Milky Way. Its access to the Southern sky will allow us to obtain more precise 3D parameters of the Galactic bar, reveal the nature of the 3-kpc Arm, and clarify the dynamical coupling between the bar and the spiral arms. By leveraging high-precision astrometry of numerous celestial objects with SKA-VLBI, the Galactic fundamental parameters such as the Solar motion and the Galactic rotation curve can be constrained more precisely. These advancements will not only elucidate the structure of our Milky Way, but also provide benchmarks for understanding barred spiral galaxies in general. Furthermore, they are important for advancing our knowledge of cosmological structure formation. The capabilities of SKA-VLBI will open a new era of high precision Galactic astrometry.
}

\begin{document}
\maketitle
\newcommand{\actaa}{Acta Astron.} 
\newcommand{\araa}{Annu. Rev. Astron. Astrophys.} 
\newcommand{\aar}{Astron. Astrophys. Rev.} 
\newcommand{\ab}{Astrobiol.} 
\newcommand{\aj}{Astron. J.} 
\newcommand{\apj}{Astrophys. J.} 
\newcommand{\apjl}{Astrophys. J. Lett.} 
\newcommand{\apjs}{Astrophys. J. Suppl. Ser.} 
\newcommand{\ao}{Appl. Opt.} 
\newcommand{\apss}{Astrophys. Space Sci.} 
\newcommand{\aap}{Astron. Astrophys.} 
\newcommand{\aapr}{Astron. Astrophys. Rev.} 
\newcommand{\aaps}{Astron. Astrophys. Suppl.} 
\newcommand{\baas}{Bull. Am. Astron. Soc.} 
\newcommand{\caa}{Chinese Astron. Astrophys.} 
\newcommand{\cjaa}{Chinese J. Astron. Astrophys.} 
\newcommand{\cqg}{Class. Quantum Gravity} 
\newcommand{\gal}{Galaxies} 
\newcommand{\gca}{Geochim. Cosmochim. Acta} 
\newcommand{\icarus}{Icarus} 
\newcommand{\jcap}{J. Cosmol. Astropart. Phys.} 
\newcommand{\jgr}{J. Geophys. Res.} 
\newcommand{\jgrp}{J. Geophys. Res.: Planets} 
\newcommand{\jqsrt}{J. Quant. Spectrosc. Radiat. Transf.} 
\newcommand{\memsai}{Mem. Soc. Astron. Italiana} 
\newcommand{\mnras}{Mon. Not. R. Astron. Soc.} 
\newcommand{\nat}{Nature} 
\newcommand{\nastro}{Nat. Astron.} 
\newcommand{\ncomms}{Nat. Commun.} 
\newcommand{\nphys}{Nat. Phys.} 
\newcommand{\na}{New Astron.} 
\newcommand{\nar}{New Astron. Rev.} 
\newcommand{\physrep}{Phys. Rep.} 
\newcommand{\pra}{Phys. Rev. A} 
\newcommand{\prb}{Phys. Rev. B} 
\newcommand{\prc}{Phys. Rev. C} 
\newcommand{\prd}{Phys. Rev. D} 
\newcommand{\pre}{Phys. Rev. E} 
\newcommand{\prl}{Phys. Rev. Lett.} 
\newcommand{\psj}{Planet. Sci. J.} 
\newcommand{\planss}{Planet. Space Sci.} 
\newcommand{\pnas}{Proc. Natl Acad. Sci. USA} 
\newcommand{\procspie}{Proc. SPIE} 
\newcommand{\pasa}{Publ. Astron. Soc. Aust.} 
\newcommand{\pasj}{Publ. Astron. Soc. Jpn} 
\newcommand{\pasp}{Publ. Astron. Soc. Pac.} 
\newcommand{\rmxaa}{Rev. Mexicana Astron. Astrofis.} 
\newcommand{\sci}{Science} 
\newcommand{\sciadv}{Sci. Adv.} 
\newcommand{\solphys}{Sol. Phys.} 
\newcommand{\sovast}{Soviet Ast.} 
\newcommand{\ssr}{Space Sci. Rev.} 
\newcommand{\uni}{Universe} 
\newcommand{\prx}{Phys. Rev. X} 
\newcommand{\lrr}{Living Rev. Relativ.} %
\newcommand{\raa}{Res. Astron. Astrophys.} %
\section{Introduction}\label{sec:introduction}

\subsection{Background}\label{sec:introduction-background}

The Milky Way is not only the only known galaxy harboring intelligent life but also one of billions of galaxies in the universe. In 2021, \textit{Science} magazine identified the question ``Is our Milky Way Galaxy special?'' as one of the 125 most pressing scientific challenges. Addressing this issue is essential not only for determining whether the Milky Way is atypical in its ability to host life but also for uncovering fundamental processes. 

\begin{figure}[!ht]
    \centering
    \includegraphics[width=0.85\linewidth]{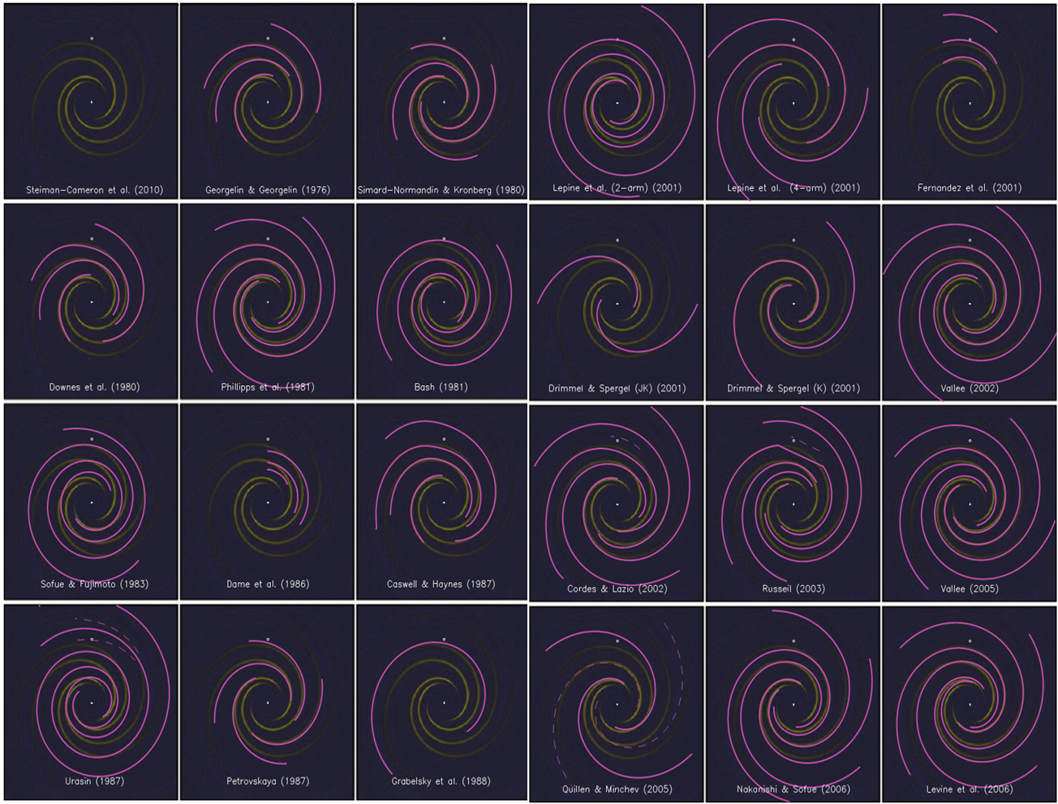}
    \caption{The spiral model from the literature \citep{Steiman+2010}.}
    \label{fig:models}
\end{figure}

The scientific study of the Milky Way has evolved over four centuries, yet its global structure remains only roughly constrained. Despite hundreds of proposed spiral arm models (see Figure~\ref{fig:models}), persistent uncertainties arise from three key observational challenges. First, our position within the Galactic disk imposes an inside-looking-out perspective, requiring observations through dense interstellar material. This leads to severe line-of-sight projection effects, as star-forming regions, molecular clouds, and dust at different distances overlap along sightlines. Second, the absorption and scattering of short-wavelength radiation by interstellar dust and gas cause optical observations to diminish significantly with distance. As such, they are ineffective for detecting stellar distributions and kinematics in the distant Galactic disk, thereby severely constraining the ability to resolve the Galactic structure. Third, extant surveys exhibit strong sampling biases: tracer objects are predominantly concentrated in the northern sky, while systematic coverage of southern sky remains critically deficient, particularly the Galactic Center region, the far side of the Milky Way and the fourth quadrant. Current models rely heavily on assumptions and extrapolations. Consequently, precise determination of the global structure of the Milky Way remains one of the most challenging tasks in modern astrophysics. 

\subsection{Motivation}\label{sec:introduction-motivation}

Each advancement in observational capabilities has revolutionized our understanding of the Milky Way. As the most powerful radio interferometer to be constructed, SKA-VLBI will resolve long-standing questions in Galactic structure through its unprecedented sensitivity and unique southern hemisphere perspective.

SKA-VLBI can provide superior parallax accuracy relative to \textit{Gaia} and is unaffected by interstellar dust extinction, unlike optical observations, making it particularly advantageous for precise and accurate astrometry in the Galactic disk. With the SKA-VLBI array we can aim at the first complete 3D mapping of the Galactic structure. Located predominantly in the southern hemisphere, the SKA-VLBI will provide complementary coverage to northern surveys such as the BeSSeL \citep{Brunthaler+2011} and VERA \citep{VERA+2020} programs, thereby eliminating current hemispheric sampling biases and providing a unified view of the Galactic structure. SKA-VLBI will: (1) detect faint tracers to accurately constrain the morphology of local spiral arms, and (2) detect distant spiral features essential for a complete Galactic reconstruction. (3) The SKA-VLBI's unique coverage of the Galactic Center will allow us to investigate the connection between the Galactic bar and spiral arms, constraining their dynamical interactions. 

By delivering precise trigonometric parallax and kinematic measurements of star-forming regions across the disk, SKA-VLBI will enable more accurate determinations of the Galactic fundamental parameters and also provide critical insights into the kinematic properties of spiral arms.

Various maser species lie in the frequency range of SKA-VLBI that can be used for astrometry, in particular class II methanol at 6.7 and 12.2 \,GHz and 22\,GHz water masers \citep[for more information about SKA-Mid maser science of HMSFRs, please refer to SKA Maser chapter from Our Galaxy][]{Rygl+2026}. These masers arise in HMSFRs (methanol, water) and evolved stars (water), making them effective probes of star formation processes. 
Furthermore, their precise positions and kinematics provide valuable constraints for calibrating kinematic distances \citep[e.g., BeSSeL kinematic distance calculator][]{Reid+2022}.

\section{Specifications of SKA-VLBI}

According to the recent review by \citet{Li+2024}, VLBI with SKA-Mid is expected to achieve a typical baseline of $\sim$10,000~km. Based on the BeSSeL survey, the majority of published targets consists of 6.7 and 12.2~GHz methanol masers, as well as 22~GHz water masers. Given the frequency range of the SKA, it will observe the 6.7 and 12.2~GHz methanol masers first. Due to ionospheric effects, 12.2~GHz methanol masers are better suited for tracing distant structures, while 6.7~GHz methanol masers can be applied to densify the mapping of nearby spiral arms. \cite{Hyland+2025} proposed a 6.7 GHz methanol maser survey using the Australian Long Baseline Array (LBA), suggesting that the more compact masers are likely associated with nearby spiral arms, whereas less compact masers tend to be associated with distant arms or central Galactic positions. Furthermore, the unprecedented sensitivity of the SKA when employed as a VLBI element will be crucial for enabling large-scale astrometric measurements of 6.7 GHz methanol masers. Therefore, these measurements help trace the primary skeleton of the Galactic spiral structure. Starlink poses a potential risk of radio frequency interference (RFI) to observations at 12.2~GHz, as its Ku‑band downlink frequencies overlap with this spectral window. Furthermore, SKA-VLBI could also obtain precise distances to the many water masers in the Milky Way if its frequency coverage were to be extended. The compactness of methanol and water masers would enable the long-baseline advantages offered by SKA-VLBI. At 6.7~GHz, the typical system equivalent flux density (SEFD) for SKA-VLBI is $\sim$3.8~Jy, where the SEFD for AA4 of SKA-Mid is $\sim$2.6~Jy\footnote{Additional details available via the EVN calculator: \url{https://services.jive.eu/evn-calculator/cgi-bin/EVNcalc.pl}}. Similarly, at 12.2~GHz, we assume an SEFD of $\sim$9.7~Jy \citep{Li+2024}. The corresponding theoretical sensitivity ($\Delta S$), without considering the sampling effect, is summarized in Table \ref{tab:SKA-VLBI-specification}.

The distances to masers are measured using two primary calibration techniques: phase-referencing \citep{Reid+2009a} and MultiView \citep{Rioja+2017}. Phase-referencing, a well-established method, was extensively employed in the BeSSeL survey. The MultiView technique offers the potential to `universal' high-precision astrometry, including in the low-frequency regime dominated by ionospheric disturbances, where other phase-referencing methods result in large errors \citep{Rioja+2020}. Its principal advantages are that it is not constrained by the angular separation between the calibrator and the target source. Moreover, it can achieve more accurate measurements of the systematic offset and approach the theoretical noise limit.

Based on the calculations in \citet{Li+2024} and adopting a Dynamic Range (DR) of 100:1, we have calculated several key parameters for SKA-VLBI observations, such as the position measurement uncertainty ($\Delta \theta$). For our analysis, we assume that the parallax precision to be half of the astrometric positional uncertainty. The complete results are presented in Table~\ref{tab:SKA-VLBI-specification}.

The MultiView technique offers an effective strategy for measuring maser parallaxes. To achieve optimal UV coverage, a six-hour observing epoch is recommended. Owing to the unprecedented sensitivity of the SKA, approximately 25 sources can be observed within a single epoch. Parallax measurements require six such epochs, corresponding to roughly 1.5 hours of total integration time per source (including the associated calibrator observations and slewing). Currently, trigonometric parallaxes have been measured for about 200 masers. Using the SKA, doubling this sample would require only about 300 hours of observing time—yielding results comparable to those of the BeSSeL project, which accumulated several thousand hours of observations.

\begin{table}[htbp]
	\centering  
	\setlength\tabcolsep{16pt}
	\begin{threeparttable}
		\caption{Characterizing SKA-VLBI Astrometry for Methanol Masers \label{tab:SKA-VLBI-specification}}
		\begin{tabular}{cccccc}
			\hline \hline
			Frequency & Bandwidth & SEFD & $\Delta S$  & $\theta_{\mathrm{beam}}$ & $\Delta \theta$ \\
			(GHz) & (Hz) & (Jy) & ($\mu$Jy beam$^{-1}$) & (mas) & ($\mu$as) \\
			\hline
			6.7  & $256 \times 10^{6}$  & 3.8 &	2.8 & 0.9 & 5 \\
			6.7  & $2 \times 10^{3}$  & 3.8 &	1000   & 0.9 & 5 \\	
			12.2 & $256 \times 10^{6}$	& 9.7 &	7.2 & 0.5 & 3   \\
            12.2 & $2 \times 10^{3}$  & 9.7 &	 2600 & 0.5 & 3  \\	
			\hline
		\end{tabular}
		\begin{tablenotes}
			\item{\textbf{Note:}} Assuming the baseline length is 10,000 km and the typical telescope aperture is 32 m. For $\Delta S$, the integration time is 1 hr.
		\end{tablenotes}
	\end{threeparttable}
\end{table}

\section{Science cases}\label{sec:science}
\subsection{The whole spiral structure of the Milky Way}\label{subsec:whole_spiral}

The exploration of the large-scale structure of the Milky Way has demonstrated remarkable synergy between observational advances and theoretical modeling. Radio tracers unaffected by dust extinction provide particularly valuable probes for mapping the global structure of the Milky Way. \cite{Oort+1958} established the first large-scale structure of our Milky Way by combining HI observations from both hemispheres, tracing features extending to tens of kiloparsecs. Recent studies using atomic and molecular gas indicate that the northern Galactic spiral arms may reach $\sim$22~kpc from the Galactic Center \citep{Sun+2024}. These works, however, relied on kinematic distances, which are inherently limited by uncertainties in the Galactic rotation model and the presence of non-circular (peculiar) motions. 

In the early 21st century, the advent of VLBI trigonometric parallax measurements marked the beginning of high-precision studies of Galactic structure \citep{Xu+2006}. Through a synthesis of high-precision astrometric data from the BeSSeL and VERA programs, \citet{Reid+2019} obtained precise distance and proper motion measurements for approximately 200 HMSFR masers. By combining these direct distance measurements with constraints derived from spiral arm tangent points, they constructed an improved model of the Galactic spiral structure extending to galactocentric distances of $\sim$15~kpc. \citet{Reid+2019} represents the most accurate and comprehensive 3D reconstruction of the Galactic structure achieved to date through geometric distance measurements.

While the spiral structure of the Milky Way constructed from VLBI trigonometric parallax measurements marks significant progress, it still faces several key limitations. First, the severe lack of precise distance measurements in the fourth Galactic quadrant and the region towards and beyond the Galactic Center. Currently, only one HMSFR maser with a measured trigonometric parallax exists beyond the Galactic Center at distances about 20~kpc from the Sun \citep{Sanna+2017}. Consequently, investigations of the spiral arm morphology in these regions rely on extrapolations from the northern arm segments and tangent-point constraints. Second, the geometric distance model ($\sim$15~kpc) exhibits a more limited spatial coverage compared to atomic/molecular gas tracers ($\sim$22~kpc). These limitations arise primarily from the predominantly northern hemisphere distribution of VLBI facilities and the sensitivity constraints of current surveys. These challenges can be effectively resolved by SKA-VLBI.

\begin{figure}[ht]
	\centering
	\includegraphics[width=0.7\columnwidth]{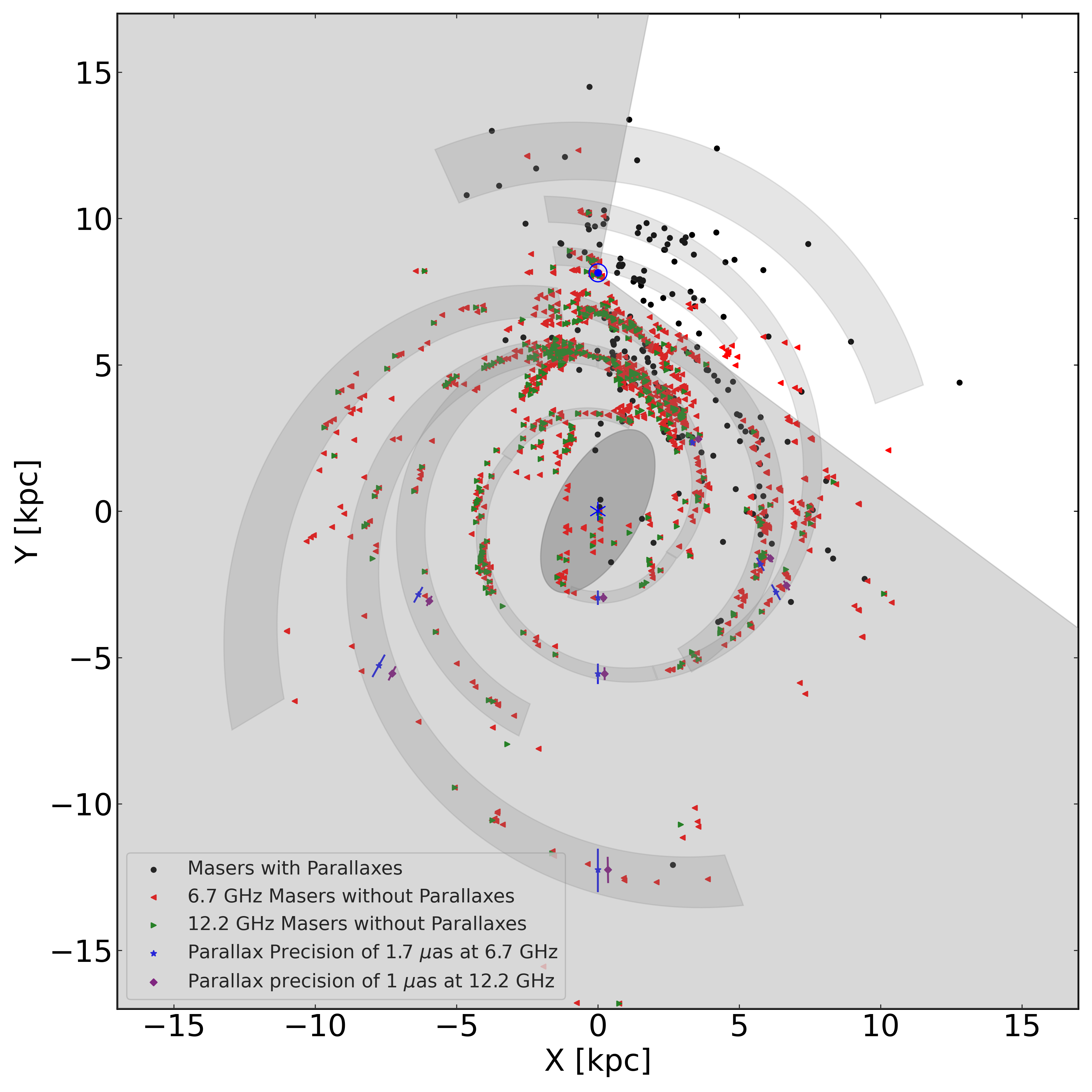}
	\caption{Distribution of masers in the Milky Way. Black dots show the maser sample compiled by \citet{Reid+2019}. Red and green triangles indicate the known 6.7 and 12.2~GHz methanol masers detectable with SKA-VLBI, respectively. The full references for these masers are compiled in \cite{Li+2024}. The distance is estimated using the kinematic method. Blue and purple diamonds represent simulated sources with with assumed parallax precisions of 1.7 and 1~$\mu$as, respectively. The gray shaded region shows the coverage area of SKA-VLBI assuming Galactic latitude of 0$^{\circ}$ \citep{Li+2024}. The ellipse indicates the bar. The Galactic Center (blue star) is at (0, 0)~kpc and the Sun (blue Sun symbol) is at (0, 8.15)~kpc.}
	\label{fig:masers}
\end{figure}

Figure~\ref{fig:masers} presents the distribution of 6.7~GHz (1875) and 12.2~GHz (433) methanol masers detectable by SKA-VLBI in the Milky Way. The SKA-VLBI will provide complete coverage of the fourth Galactic quadrant for the first time, addressing the current critical shortage of maser astrometric data in this region.  With its theoretical parallax precision of 1.7~$\mu$as, i.e., 1/3 of the astrometric precision shown in Table~\ref{tab:SKA-VLBI-specification} \citep{Li+2024}, SKA-VLBI will enable both precise distance determinations of these masers, facilitating direct 3D mapping of spiral arm structures. In addition, SKA-MID (not VLBI) can be used to measure 3D kinematic distances, a particularly effective method for objects on the far side of the Milky Way (\citealt{Reid+2022}, and SKA Maser chapter from Our Galaxy \citealt{Rygl+2026}). When combined with existing northern sky surveys, these data will provide the first model-independent, geometric reconstruction of the global structure of the Milky Way.

Moreover, SKA-VLBI will provide critical constraints on spiral arm structures beyond the Galactic Center, eliminating current reliance on purely logarithmic spiral extrapolations. Assuming a spiral arm width of 0.3~kpc \citep{Reid+2019}, the astrometric precision of SKA-VLBI remains sufficient to unambiguously determine spiral arm associations for masers out to 17~kpc. This will not only extend the reliably mapped extent of the Milky Way but also provide robust constraints on the true radial extent of the spiral arms. These capabilities will revolutionize our understanding of the spiral structure of our Milky Way, potentially resolving debates about the Galactic structure that have persisted for over half a century.

\subsection{The inner part structure of the Milky Way}\label{sec:spiral-sturcture-whole}

The inner Galactic region represents a crucial area for studying the structure and dynamics of the Milky Way. It hosts the launching and bifurcation points of spiral arms, the Galactic bar, and the 3-kpc Arm \citep{Bland+2016}. Despite its importance, our understanding of this region remains incomplete due to the combined effects of heavy extinction and crowding, which present significant observational challenges. The predominance of telescopes in the northern hemisphere exacerbates this problem, since they cannot fully cover the inner region of the Milky Way. These factors collectively result in a critical shortage of precise observational data for this region.

\begin{figure}[ht]
	\centering
	\includegraphics[width=0.45\columnwidth]{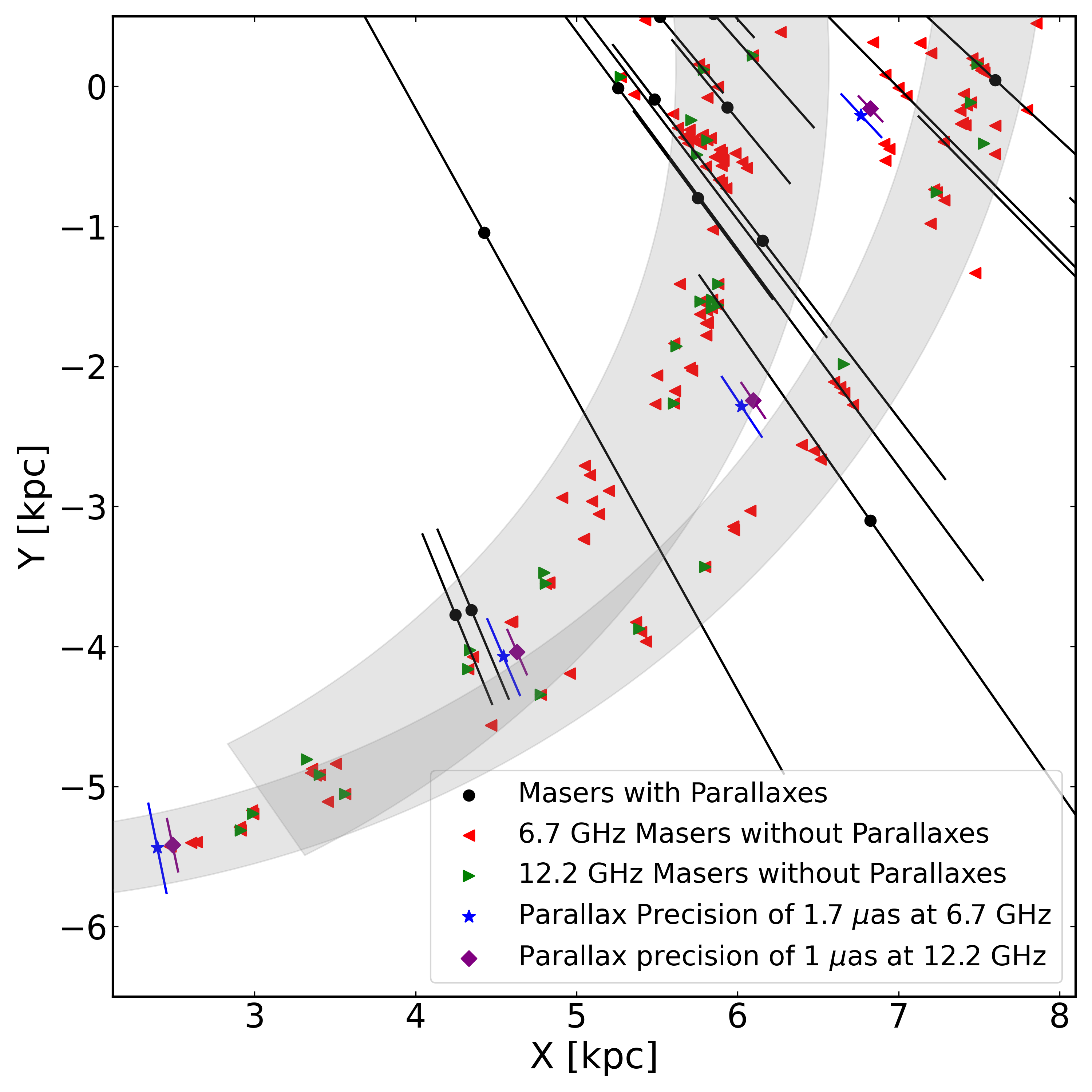}
	\includegraphics[width=0.45\columnwidth]{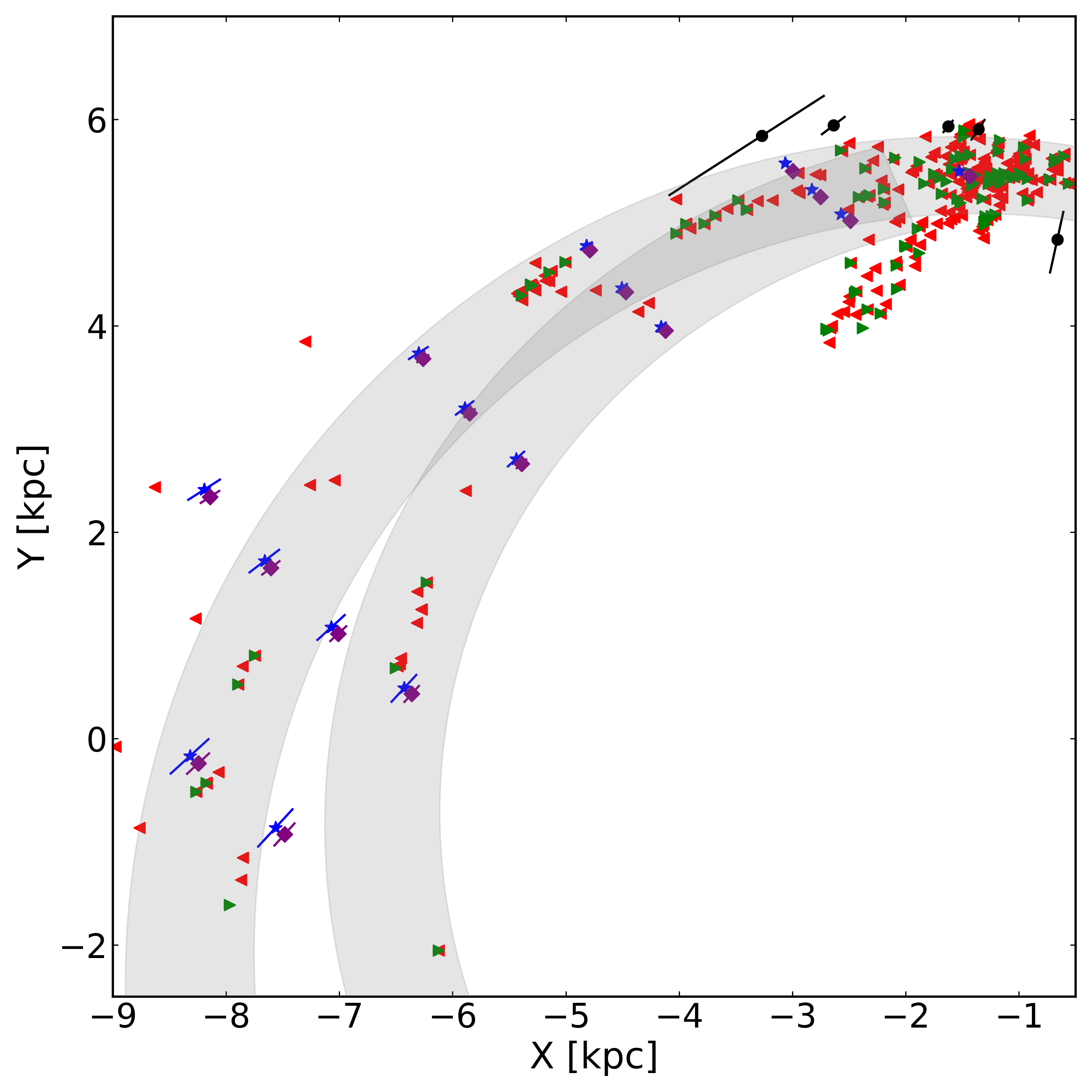}
	\caption{HMSFR masers near spiral arm bifurcation points \citep{Xu+2023}. The left panel illustrates the bifurcation between the Sagittarius and Perseus arms in the first Galactic quadrant, and the right panel displays the bifurcation between the Centaurus and Norma arms in the fourth Galactic quadrant. All other elements follow the same convention as in Figure~\ref{fig:masers}.}
	\label{fig:bifurcation-maser-all}
\end{figure}

Using the same sample of HMSFR masers, \citet{Reid+2019} and \citet{Xu+2023} have proposed competing models for the inner spiral structure of the Milky Way. \citet{Reid+2019} favor a four-armed spiral pattern, consistent with the prevailing paradigm. \citet{Xu+2023} advocate for a two-armed inner structure accompanied by multiple outer arms, a morphological structure that is more commonly observed in external spiral galaxies \citep{Wei+2024}. The bifurcation points of spiral arms serve as critical regions for discriminating between these models. As illustrated in Section~\ref{subsec:whole_spiral}, SKA-VLBI will be able to trace the spiral arms out to 17~kpc, thereby covering the predicted bifurcation regions outlined by \citet{Xu+2023} (see Figure~\ref{fig:masers}). Hence, it will provide direct constraints on spiral arm morphology (see Figure~\ref{fig:bifurcation-maser-all}) and reveal how spiral arms connect at bifurcation points.

The Galactic bar is predominantly composed of stars rather than gas or dust, making red clump giants (a subcategory of evolved stars) the most effective tracers of its structure \citep{Wegg+2015}. However, severe interstellar extinction remains a significant obstacle to obtaining precise structural parameters of the bar. Notably, the rotation of the bar perturbs the surrounding interstellar medium, triggering localized star formation. HMSFR masers can be used to probe the outer boundary of the bar. Since radio observations are unaffected by dust extinction, they offer a new and independent means to constrain the properties of the bar. \citet{Kumar+2025} fitted the 3D structure of the bar using 15 HMSFR masers located within 4~kpc of the Galactic Center. Their best-fit model yields a bar semi-major axis of $4.0 \pm 0.1$~kpc, semi-minor axis of $1.9 \pm 0.2$~kpc, and a Galactocentric azimuth of $36^{\circ} \pm 5^{\circ}$. Future observations with SKA-VLBI are expected to further refine these parameters. As mentioned in Section~\ref{subsec:whole_spiral}, there are over 200 methanol masers (at 6.7 or 12.2~GHz) located within 4~kpc of the Galactic Center. This large sample will enable statistically robust, high-precision studies of the morphology and dynamics of the Galactic bar in unprecedented detail.

High-precision kinematic observations of the Galactic bar ends provide crucial insight into the dynamical coupling between the bar and spiral arms. These measurements offer direct tests of the prevailing theory that spiral arms are driven by the Galactic bar. Recent studies by \citet{Immer+2019} and \citet{Li+2022} have found that the Galactic bar introduces significant non-circular motions in spiral arms near the bar-end region. Their conclusions are based on the 3D kinematic analysis of 16 HMSFR masers distributed across the first Galactic quadrant, tracing both the bar end and the adjacent Scutum arm. As shown in Figure~\ref{fig:masers}, a rich population of masers are located near the bar end in the first Galactic quadrant. SKA-VLBI will provide complete spatial coverage of this region. Microarcsecond level astrometry of these maser sources will yield powerful new constraints on spiral arm formation mechanisms, offering fresh perspectives on Galactic dynamics.

The nature of the 3-kpc Arm has been debated since it was discovered in 1957 \citep{Woerden+1957}. Recently, \citet{Kumar+2025} analyzed VLBI astrometry and 3D kinematics of 14 HMSFR masers associated the 3-kpc Arm, offering new insights into this long-debated structure. Their results indicate that the so-called ``Expanding 3-kpc Arm'' is instead consistent with quasi-elliptical orbits driven by the Galactic bar. They suggest that the stars and gas identified with the 3-kpc Arm are neither expanding nor maintaining a fixed 3 kpc radius, nor exhibiting spiral arm characteristics. However, these conclusions currently rely on a limited sample of 14 HMSFR masers. As shown in Figure~\ref{fig:masers}, future SKA-VLBI observations will revolutionize our understanding of this region through their unparalleled sensitivity and angular resolution. By detecting a larger population of masers and obtaining precise distance and kinematic measurements, it will provide definitive evidence to determine whether the 3-kpc Arm constitutes an independent structure or represents a bar-driven dynamical feature.

\subsection{The Fundamental Parameters of the Milky Way}\label{sec:spiral-sturcture-parameter}

The Milky Way represents the only spiral galaxy that we can conduct detailed investigations. However, persistent uncertainties in Galactic fundamental parameters make it hard to estimate the total mass (including the dark matter halo) of the Milky Way, accurately transform from Heliocentric to Galactocentric coordinate systems for modeling the Local Group dynamics, and make robust predictions of dark matter annihilation radiation \citep{Reid+2014b}. Resolving these uncertainties is critical for developing a coherent understanding of the structure and evolution of the Milky Way.

The vertical position of the Sun with respect to the Galactic plane ($z_{\odot}$) remains debated, with estimates varying between 20 and 30~pc. Kinematic parameters show comparable uncertainties. The solar motion components ($U_{\odot}, V_{\odot}, W_{\odot}$) exhibit $>$10~km~s$^{-1}$ discrepancies, particularly in $V_{\odot}$. The Galactic rotation curve remains poorly constrained in both the inner ($R < 4$~kpc) and outer ($R > 12$~kpc) regions. See \citet{Bland+2016} for a comprehensive review. 

Recent modeling based on HMSFR masers yields $R_{0}$ = 8.15$\pm$0.15~kpc and $\Theta_{0}$ = 236$\pm$7~km~s$^{-1}$, with a nearly flat rotation curve between 4 and 13~kpc from the Galactic Center \citep{Reid+2019}. These values were derived by fitting a model for Galactic circular rotation to the positions and velocities of the HMSFR masers. The $\Theta_{0}$ value exceeds the IAU-recommended value of 220~km~s$^{-1}$ by $\sim$10\%, corresponding to a $\sim$30\% increase in the estimated total mass of the Milky Way \citep{Reid+2014b}. However, the current sample of HMSFR masers shows spatial inhomogeneity. More than 90\% of precisely measured HMSFR masers are concentrated within the range of $0^{\circ} < \beta < 90^{\circ}$, while other regions remain severely undersampled. This uneven distribution may introduce substantial systematic uncertainties in the determination of the fundamental parameters of the Milky Way.

SKA-VLBI is expected to help to derive more accurate Galactic fundamental parameters. The unparalleled sensitivity and spatial coverage of SKA-VLBI will enable more uniform observations of these masers, spanning most Galactic regions. Based on the simulation by \citet{Quiroga+2017}, Galactic parameter estimates can be substantially improved and their error bars significantly reduced by incorporating additional maser observations, particularly those located predominantly in the southern hemisphere. This improved sampling will enhance the reliability and precision of kinematic model fitting, thereby reducing uncertainties in the derived parameters. 

\section{Summary and Conclusion}\label{sec:summary}

This paper highlights the transformative potential of next-generation SKA-VLBI observations for Milky Way studies. With its unparalleled sensitivity, astrometric precision, and strategic  southern hemisphere location, SKA-VLBI will enable 3D mapping of Galactic spiral structure spanning from the solar neighborhood to the Galactic Center and to the far side of the Galaxy. These observations will resolve long-standing debates regarding the number, morphology, and spatial distribution of spiral arms, particularly in the inner regions of the Milky Way, where current models exhibit discrepancies. 

Additionally, SKA-VLBI will constrain the structural parameters of the Galactic bar and probe the nature of the 3-kpc Arm. By measuring precise positions and kinematics for numerous HMSFR masers, the SKA-VLBI will refine estimates of the Galactic fundamental parameters, providing tighter constraints on both the mass distribution and dark matter halo profile of the Milky Way. 

These advancements will not only improve our understanding of the Milky Way but also establish new benchmarks for Galactic astronomy. The SKA-VLBI will usher in a new era of high precision Galactic astronomy. 

\textbf{\large Acknowledgements} 

This work is supported by the National SKA Program of China (grant No. 2022SKA0120103), the NSFC Grants Nos. 12203104, 12403077, 12403041 and 12503071, and the Key Laboratory for Radio Astronomy. 

\bibliographystyle{abbrvnat-maxbibnames4}
\bibliography{Spiral} 

\end{document}